\documentclass{article}

\usepackage{arxiv}

\usepackage[utf8]{inputenc} % allow utf-8 input
\usepackage[T1]{fontenc}    % use 8-bit T1 fonts
\usepackage{hyperref}       % hyperlinks
\usepackage{url}            % simple URL typesetting
\usepackage{booktabs}       % professional-quality tables
\usepackage{amsfonts}       % blackboard math symbols
\usepackage{nicefrac}       % compact symbols for 1/2, etc.
\usepackage{microtype}      % microtypography
\usepackage{lipsum}
\usepackage[font=small,labelfont=bf]{caption} 

\usepackage{graphicx,calc}
\usepackage{amsmath}
\usepackage[english]{babel}
\usepackage{eurosym}
\usepackage{stmaryrd}

\usepackage{comment}

\graphicspath{{IMGs//}}

\title{Bimodal buckling governs human fingers' luxation}

\author{
 Massimiliano Fraldi\\
  Department of Structures for Engineering and Architecture, and \\
  Laboratory of Integrated Mechanics and Imaging for Testing and Simulation (LIMITS),\\ University of Napoli "Federico II", 80125 Napoli, Italia\\
  \texttt{fraldi@unina.it} \\
  \And
  Stefania Palumbo \\
Department of Structures for Engineering and Architecture, and \\
  Laboratory of Integrated Mechanics and Imaging for Testing and Simulation (LIMITS),\\ University of Napoli "Federico II", 80125 Napoli, Italia\\
  \texttt{stefania.palumbo@unitn.it} \\
  \And
  Arsenio Cutolo \\
  Department of Structures for Engineering and Architecture, and \\
  Laboratory of Integrated Mechanics and Imaging for Testing and Simulation (LIMITS),\\ University of Napoli "Federico II", 80125 Napoli, Italia\\
  \texttt{arsenio.cutolo@unina.it}\\
  \And
   Angelo Rosario Carotenuto\\
  Department of Structures for Engineering and Architecture, and \\
  Laboratory of Integrated Mechanics and Imaging for Testing and Simulation (LIMITS),\\ University of Napoli "Federico II", 80125 Napoli, Italia\\
   \texttt{angelorosario.carotenuto@unina.it}\\
   \And
  Davide Bigoni\\
  Department of Civil, Environmental and Mechanical Engineering, University of Trento, 38123 Trento, Italia\\
  \texttt{davide.bigoni@unitn.it} 
}

\begin{document}
\maketitle

\begin{abstract}
Equilibrium bifurcation in natural systems can sometimes be explained as a route to stress shielding for preventing failure. Although compressive buckling has been known for a long time, its less-intuitive \textit{tensile} counterpart was only recently discovered and yet never identified in living structures or organisms. Through the analysis of an unprecedented \textit{all-in-one} paradigm of elastic instability, it is theoretically and experimentally shown that coexistence of two curvatures in human finger joints is the result of an optimal design by nature that exploits both compressive and tensile buckling for inducing luxation in case of traumas, so realizing a unique mechanism for protecting tissues and preventing more severe damage under extreme loads. Our findings might pave the way to conceive complex architectured and bio-inspired materials, as well as new artificial joint prostheses and robotic arms for bio-engineering and healthcare applications. 
\end{abstract}

% keywords can be removed

\section{Introduction}
\begin{figure*}%[!h] %htbp
\centering
\includegraphics[width=1\textwidth]{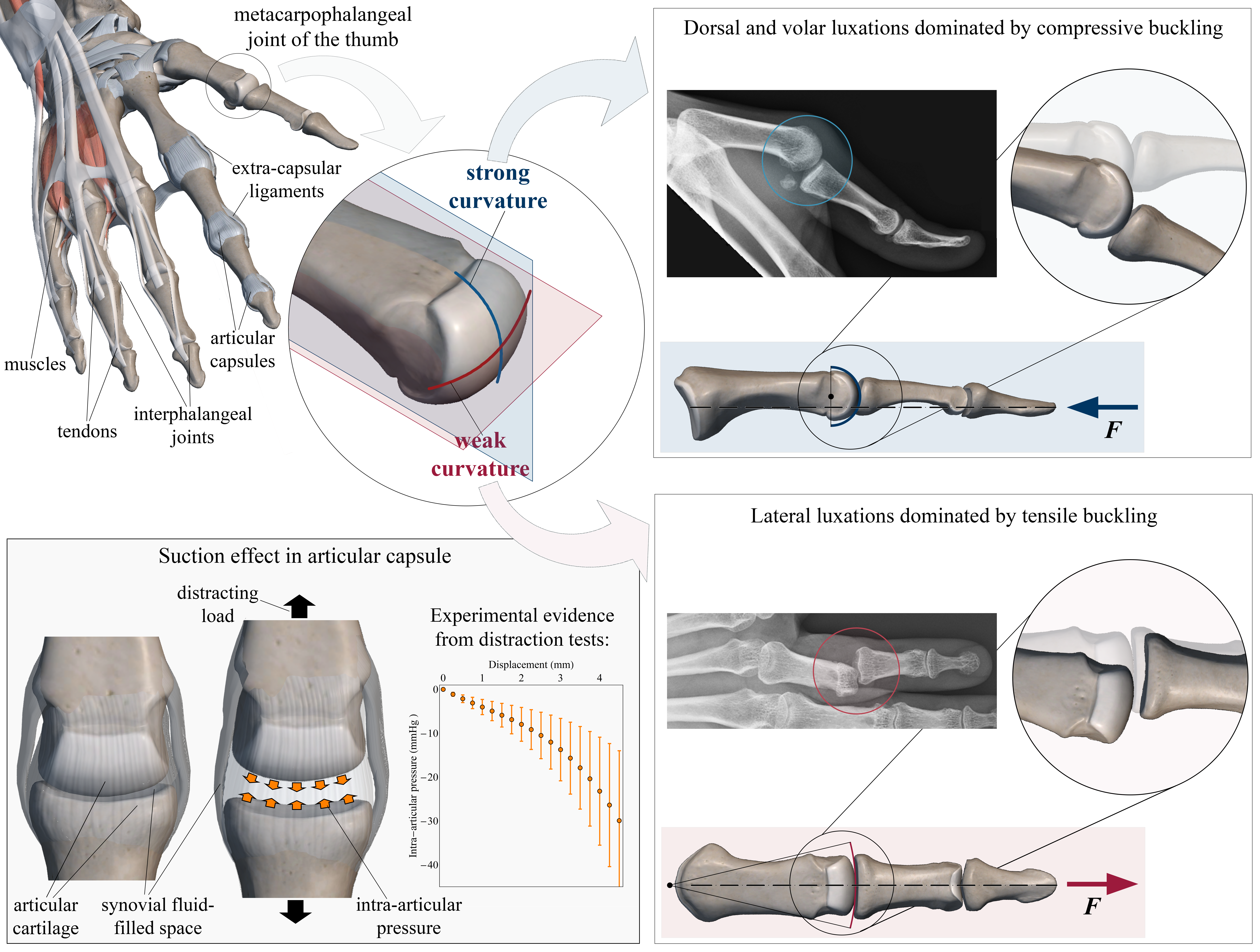}
\caption{On the left, a sketch of the anatomy of the human hand highlighting the main components of the musculoskeletal system that drive the mechanics of finger joints, with a focus on the suction-like effect limiting articular distractions, inset adapted from \cite{NIPhand1}. On the right, sketches and original RX images reproducing the complementary mechanisms of finger joints' dislocation occurring in orthogonal anatomical planes of the hand: example of compression-induced dorsal/volar luxation in the sagittal (or lateral) plane, characterized by a \textit{strong} curvature of the bone terminals, and illustration of a lateral luxation determined by a high tensile load in the transversal (horizontal) plane, where a \textit{weak} curvature marks the geometry of the bones' ends.}
\label{fig.Figure2}
\end{figure*}

Bones are connected through joints, the articulations, permitting the specialized movements necessary for everyday activities and, simultaneously,  providing stability to the musculoskeletal system. \\
Among the structurally and functionally different types of articulations, synovial joints --also known as diarthroses-- offer the highest degree of motion and are typical of shoulders, knees, elbows, hips, hands, and feet \cite{graysanatomy,guidugli}, which are all indeed implicated in locomotion and handling of objects \cite{Gray}.\\
In human fingers, the diarthroses house the terminals of the joined bones in a cavity enveloped by an articular capsule \cite{ralphs1994} that structurally connects the adjacent bones by way of its outermost fibrous layer and secretes a viscous fluid filling the space of the cavity through an inner membrane, the synovium, Fig. \ref{fig.Figure2}. Both the synovial fluid and the sheets of hyaline cartilage that coat bones' articulating surfaces eliminate friction and absorb shocks during movements \cite{Mow1993,neville2007,tissuemechanics}. Finally, extra-capsular ligaments \cite{fungbook,tissuemechanics} and the suction-like effect, which provides negative intra-articular pressure in response to bone ends' distraction (i.e. separation)  \cite{NIPhand1,NIPshoulder1,tribon,NIPshoulder2,shoulderstab1,NIPshoulder3},  contribute towards mechanical stabilisation of the joint \cite{fingerstab}.\\
As a result of their continuous involvement in body movements, synovial joints are frequently subject to injuries, luxations being among the most common ones. They consist in an abnormal displacement between the articulating bones, whose  ends in contact move out of their anatomical position, either for returning back (sub-luxation) or fully and irreversibly dislocating (complete luxation) \cite{PIPjoints,fingerdisloc,SUNDARAM2013}. In fingers, dislocations typically occur in case of impacts due to falls or collisions during sport \cite{elzinga2017finger,basketball}, or as a consequence of over-stretching caused by climbing or accidental fingers' trapping \cite{bach1999finger,logan2004,gnecchi2015}.\\
From a mechanical standpoint, luxations can be seen as the response of the bone-joint-bone structural system to either abnormal compressive or tensile forces, giving rise to the two mechanisms sketched in Fig. \ref{fig.Figure2} and highlighted by X-ray images as well. The geometrical configurations assumed by dislocated fingers, which recall deviation of hinged bars due to elastic stability, as well as the extremely regular shape of the bone epiphyses shown in Fig. \ref{fig.Figure2}, prompts the question of whether luxations may conceal a mechanical strategy to preserve bone integrity and minimize irreversible tissues damages, in the event of extraordinary loads. \\
To investigate this theoretically, and  inspired  from the two different curvatures exhibited by the bone ends at the joint level, a new structural \textit{all-in-one} paradigm is introduced in the present article, capable of undergoing both tensile and compressive buckling in two orthogonal planes, whose resulting kinematics  retraces the corresponding two above mentioned dislocations phenomena. To best mimic the real physiology of the finger's joint, the  mechanical model is  equipped with elastic elements simulating ligaments and the suction effect due to the synovial capsule is additionally incorporated. \\
In this way, buckling-induced deviation of bone segments from their physiological configuration is demonstrated to allow a strong decrease of the overall elastic energy and a relief of mechanical stresses, so preventing cracking in bone and tears of ligaments and tendons, under both compressive and tensile accidental loads. Our conclusion is that the finger structure not only provides the first example in nature of tensile buckling \cite{zaccaria2011}, but might also represent a unique biological system whose geometrical and mechanical features are optimally designed to shield tissues from high stresses, by harnessing in one system both compressive and tensile elastic instabilities \cite{elasticstability,Bigoni2012,1dsd,paradox}. This could open new perspectives for integrated neuro-mechanical design of bionic hands, robotic prostheses and exoskeletons for human rehabilitation and other bioengineering applications \cite{siviy2022}.

\section{A luxation-inspired \textit{all-in-one} buckling paradigm}
\begin{figure*}[!h] %htbp
\centering
\includegraphics[width=1\textwidth]{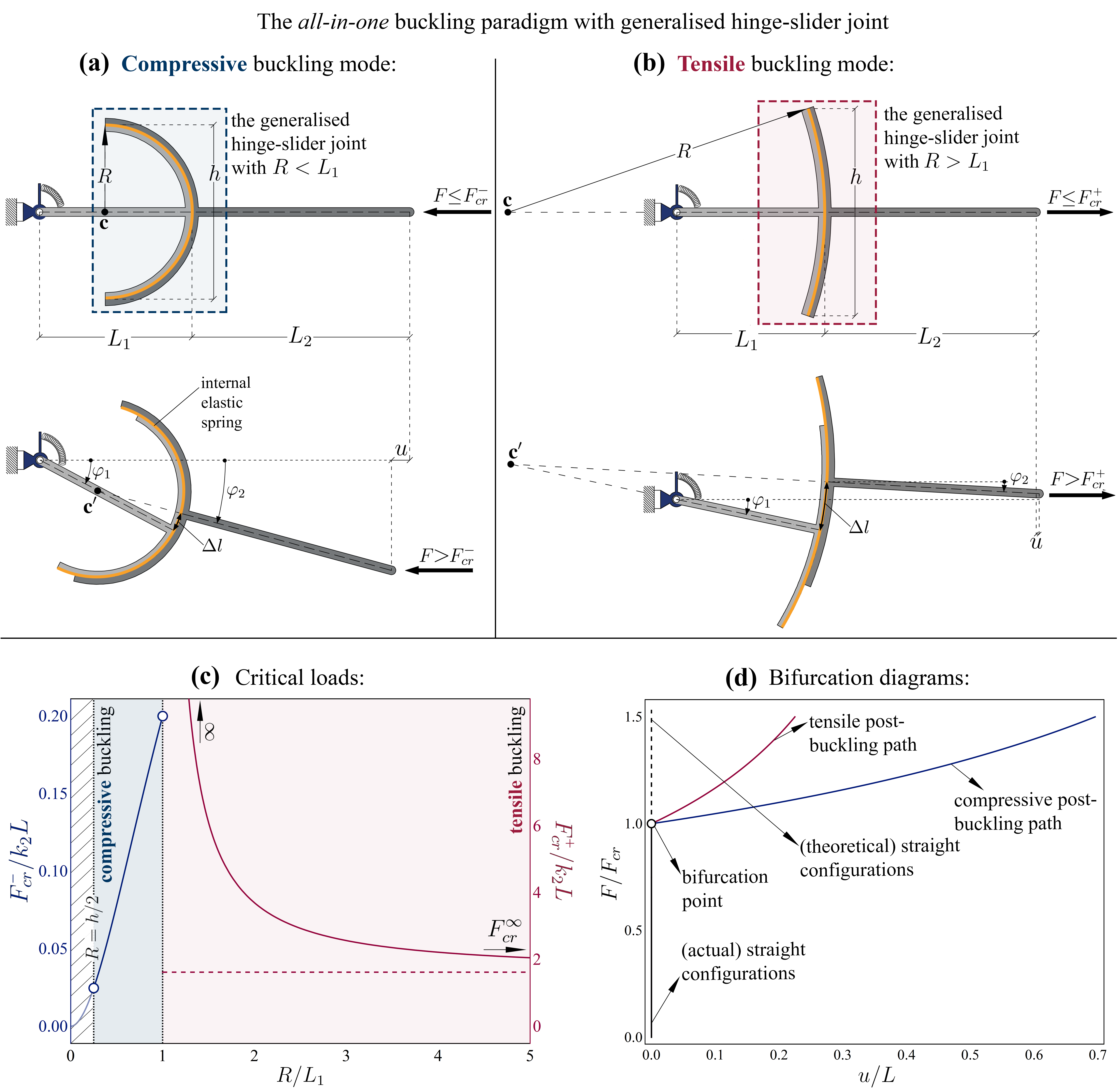}
\caption{Structural scheme of the proposed \textit{all-in-one} paradigm of elastic instability undergoing \textbf{(a)} compressive and \textbf{(b)} tensile buckling in orthogonal planes as a function of the ratio between the joint's radius and the hinged rod's length. \textbf{(c)} (Normalised) critical load \textit{versus} ratio between the joint's radius and the hinged rod's length, for an illustrative case with $L_1=L_2=L/2$, $h=L/2$ and $k_1=k_2 L^2$. Herein, compressive ($h/2\leq R < L_1$) and tensile ($R>L_1$) buckling domains are evidenced, along with the geometrically incompatible region where $R<h/2$. Also, the divergence of the tensile critical load for $R/L_1 \rightarrow 1$ is highlighted as well as its asymptotic value $F_{cr}^{\infty}=k_2 L/2 +\sqrt{k_2\left(k_2 L^2+4k_1\right)}/2$ corresponding to the limit of straight slider, i.e. $R/L_1 \rightarrow \infty$. \textbf{(d)} Examples of equilibrium bifurcation diagram for both the cases of structure buckling under axial compression (for $R=h/2=L/4$; blue curve; $F_{cr}=F_{cr}^{-}$) and tension (for $R=3L/4$; red curve; $F_{cr}=F_{cr}^{+}$), by starting from an initially stable straight configuration (solid black curve) that becomes unstable (dashed black curve) beyond the bifurcation point.}
\label{fig.Figure1}
\end{figure*}

The epiphyses of fingers' bones, at interphalangeal and metacarpophalangeal joints, evidence two remarkably different curvatures in the two orthogonal anatomical planes, which affect type and range of movements of the articulated bone segments (Fig. \ref{fig.Figure2}). In fact, an ideal longitudinal section through the sagittal plane highlights a \textit{strong} curvature allowing for smooth flexion and extension, while a \textit{weak} curvature is exhibited in the transverse plane, where mobility is almost completely locked. Apart from the role of the non-spheroidicity of the joint in limiting the physiological range of motion, we propose that the double, \lq strong and weak', curvature is additionally involved in driving fingers' luxations under severe accidental loads, as a way to prevent more serious damages of the articulation. In particular, guided by the observation of the abnormal deviation of phalanges led by luxations under both compressive and tensile extreme forces, we here interpret luxations as an elastic equilibrium bifurcation phenomenon. To this aim, an \textit{ad hoc} conceived mechanical model is introduced, equipped with a generalized slider constraint, characterized by a double curvature in two orthogonal planes (whose projections are shown in Fig. \ref{fig.Figure1}). This model allows the analysis of  the onset of elastic instability and the transition from compressive to tensile modes of buckling  in the resulting \textit{all-in-one} buckling paradigm. 
More in detail, the system comprises two rigid rods interconnected by a hinge-slider joint, which consists of two circular tracks in smooth mutual contact and only allows a relative sliding  without detachment \cite{Misseroni_2015}. The extent of sliding is elastically contrasted by an internal spring, circumferentially arranged along the tracks, and the overall equilibrium of the structure is enforced by the presence of an elastic hinge, acting at one of the ends and so realizing a cantilever configuration. Such a joint  reduces to a traditional hinge in the limit case of vanishing ratio between the radius of curvature and the rods' length \cite{paradox}, while it becomes a flat slider as in \cite{zaccaria2011} in the complementary condition of ideally infinite ratio.\\
As sketched in Fig. \ref{fig.Figure1}, when loaded through a dead axial force $F$ at its free end, the considered system can exhibit both compressive and tensile buckling, the first occurring in the plane where the radius of curvature of the joint is shorter than the hinged rod (strong curvature plane), while the second taking place in the orthogonal direction, where the centre of relative rotation between the rigid tracts falls beyond the hinge (weak curvature plane). In the figure, $\textbf{c}$ and $\textbf{c}'$ denote the rotation centers of the generalised joint in the reference and current configuration, respectively, $L_1$ and $L_2$ are the lengths of the rigid rods, while $R$ and $h$ are the radius of curvature and the transverse size of the slider, such that $h\leq 2R$. When sufficiently high, the external load induces buckling through the non-trivial deformation modes shown in Figs. \ref{fig.Figure1}\hyperref[fig.Figure1]{a} and \ref{fig.Figure1}\hyperref[fig.Figure1]{b} for compression and tension, respectively. In both cases, the kinematics is governed by two degrees of freedom, which can conveniently be  identified with the angles $\varphi_1$ and $\varphi_2$, describing the clockwise rotation of the rods with respect to the horizontal direction. The horizontal displacement of the right end, say $u$, can be written with reference to its absolute value as
\begin{equation}
u=\vert \left(R+L_2\right) \cos\varphi_2- \left(R-L_1\right) \cos\varphi_1 - L \vert ,
\end{equation}
where $L=L_1+L_2$, while the elongation $\Delta l$ of the spring associated to the slider is  
\begin{equation}
\Delta l = R \left( \varphi_1 - \varphi_2 \right).
\end{equation} 
Among the compatible deformation modes, equilibrium configurations follow from the stationarity of the total potential energy
\begin{equation}
\Pi= \dfrac{k_1}{2} \varphi_1^2+ \dfrac{k_2}{2}\Delta l^2- F u,
\end{equation}     
where $k_1$ is the stiffness of the rotational spring constraining the hinge and $k_2$ is the circumferential  stiffness of the spring acting along the slider, both assumed as linear elastic. Stationarity of $\Pi$ leads to the system of non-linear equations 
\begin{equation}
\label{eqsys}
\begin{cases} 
\left(1+\frac{k_1}{k_2R^2}  \right) \varphi_1-\varphi_2 - \frac{F}{k_2 R} \left(1- \frac{L_1}{R} \right) \sin\varphi_1 =0 \\ 
\varphi_1-\varphi_2 - 
\frac{F}{k_2 R} \left(1+\frac{L_2}{R} \right) \sin\varphi_2=0
\end{cases},
\end{equation}
which admits bifurcation of the equilibrium solution for both the conditions of strong and weak curvature. In particular, bifurcation loads can be derived as a non-trivial solution of the  linearised  equations \eqref{eqsys} with respect to the Lagrangian variables $\varphi_1$ and $\varphi_2$, thus yielding a quadratic equation for the axial force $F$, 
\begin{equation}
a_0 F^2 - a_1 F - a_2=0,  
\end{equation}
with coefficients $a_0$, $a_1$ and $a_2$ given by 
\begin{equation}
\begin{split}
a_0 &= \left(R-L_1\right)\left(R+L_2\right), \\   
a_1 &= k_1 \left(R+L_2\right)+k_2 R^2 L, \\
a_2 &= k_1 k_2 R^2. 
\end{split}
\end{equation}
A tensile buckling force, $F^+_{cr}$, is found, which refers to the weak curvature plane (where $R>L_1$), plus three compressive bifurcation forces, the lowest of which, occurring in the strong curvature plane (where $R<L_1$), can be identified as \lq critical', $F^-_{cr}$. This implies that the mechanical model predicts  deviations from its straight configuration once tensile and compressive loads overcome the respective critical thresholds 
\begin{equation}
F_{cr}^{\pm}=\dfrac{\sqrt{a_1^2+4a_0 a_2}\pm a_1}{2a_0},  
\end{equation}  
where $a_1^2+4a_0a_2 \geq 0$ when $L_2 \geq L_1$.\\
As illustrative case of the proposed \textit{all-in-one} buckling paradigm, the critical load is plotted in Fig. \ref{fig.Figure1}\hyperref[fig.Figure1]{(c)} as a function of the ratio between the radius of the sliding joint and the hinged rod length. It is possible to observe that the structure switches from compressive to tensile buckling when the radius $R$ exceeds the length $L_1$, $R=L_1$, thus  representing a limit condition for which tensile buckling does not occur. Furthermore, bifurcation diagrams, numerically  obtained  as solutions of equations  \eqref{eqsys}, are reported in Fig. \ref{fig.Figure1}\hyperref[fig.Figure1]{(d)} for both compressive and tensile buckling.\\
Finally, it is worth noticing that straightforward calculations allow to estimate the bending stiffness of the considered system at joint level as $k_2 R^2$. This indicates that the double curvature, characterizing the finger joints' anatomy, naturally provides a flexibility for the sagittal (strong curvature) plane higher than that characterizing the transverse (weak curvature) plane, consistently with the kind of mobility actually observed in fingers.

\section{Cooperation of elastic instability and suction effect in luxation of finger joints}
\begin{figure*}[!h] %htbp
\centering
\includegraphics[width=1\textwidth]{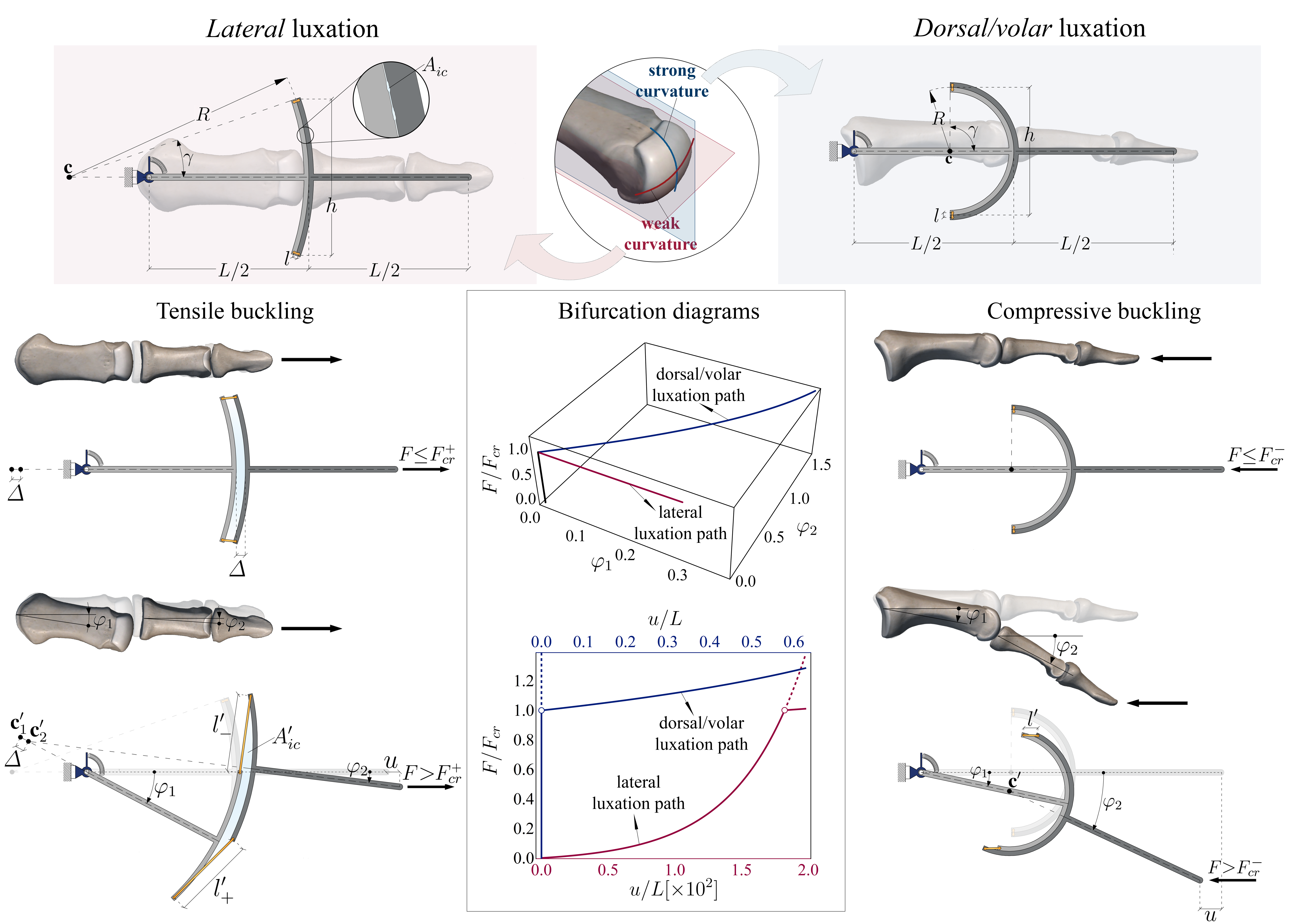}
\caption{\textbf{Left:} Structural schemes for the tensile-buckling mechanism governing lateral luxations of finger joints: the unloaded and underformed configuration (upper part), an axially distracted configuration under physiological load (center) and a buckled (laterally deviated) state induced by a load overcoming the critical threshold (lower part). \textbf{Right:} Sketch of the compressive-buckling mechanism driving finger joints' dorsal/volar luxations: the load-free configuration (upper part), the underformed configuration under physiological load (center) and a buckled (deviated) state induced by an over-critical force (lower part). \textbf{Center:} Bifurcation diagrams exhibited by the structural model at growing tensile (red curves; $F_{cr}=F_{cr}^+$) and compressive (blue curves; $F_{cr}=F_{cr}^-$) load: a 3D plot showing the dimensionless force as a function of the rotation angles (upper part) and the dimensionless force-displacement curve (lower part). Therein, solid tracts identify equilibrium paths  followed by the system, while dashed curve portions refer to theoretical unbuckled configurations. Results have been obtained by considering the following values for the geometrical and constitutive parameters: $h=0.2 L$, $s=h$, $l=0.25 L$, $A=0.1 s h$, $\beta=1$, $t=0.005 L$, $K_{ic}=0.17 E$, $k_{r}=0.001 E L^3$, with $R=0.55 L$ for the tensile case (lateral luxation) and $R=0.1L$ for the compressive buckling (dorsal/volar luxation).}
\label{fig.Figuretc}
\end{figure*}
Finger joints of human hands can be considered hallmark structures for elucidating how the unique ability to experience both compressive and tensile buckling can provide key stress shielding responses to prevent bone, ligaments and tendons from more serious mechanical damages \cite{hu2014}. To show this, the fingers' bone-joint-bone system is structurally modelled by following the above-illustrated concept of \textit{all-in-one} buckling paradigm, as reported in Fig. \ref{fig.Figuretc} with reference to the transverse and sagittal anatomical planes. Therein, to account for the stabilization role played by the capsular and extra-capsular fibrous tissues surrounding the articular joint, elastic bands connecting the ends of the slider are additionally incorporated in the mechanical model. Furthermore, anatomical considerations suggest the possibility that a small detachment between the 
epiphyses of the joined bones may occur under distracting loads. This feature is  implemented in the model and coupled with the intra-capsular suction effect that opposes the separation between articular surfaces \cite{NIPhand1,NIPshoulder1,tribon,NIPshoulder2,shoulderstab1,NIPshoulder3,ventose}. Investigation of the system’s response under uni-axial load yields the identification of 
tensile and compressive buckling as mechanisms dominating lateral and dorsal/volar luxations, respectively occurring in the transversal and sagittal planes of the hand \cite{PIPjoints}.
The progression of the dislocation process after luxation, falling beyond the scope of the present study, is neglected.

\subsection{\textit{Lateral} luxation ruled by tensile buckling along the weak curvature}
The mechanical models illustrated on the left of Fig. \ref{fig.Figuretc} show the tensile buckling mechanism through which finger joints undergo lateral dislocations in the horizontal anatomical plane, where the bones' interface is characterized by the weak curvature. In particular, starting from a resting condition, the action of a distracting force $F$ initially produces a small axial separation between the articulating bone heads, remaining within physiological limits. Then, as the magnitude of the load surpasses a critical threshold, say $F_{cr}^{+}$, the two epiphyses dislocate by deviating laterally with respect to their straight anatomical configuration. \\
The kinematic description of this bifurcation mode is characterized by three Lagrangian parameters, namely the rotations $\varphi_1$ and $\varphi_2$ of the rigid rods --representing bone segments-- with respect to the horizontal direction, and the relative displacement $\Delta$ between their ends at the joint level. Therefore, equilibrium configurations imposing stationarity of the total potential energy 
\begin{equation}
\label{PiT}
\Pi= \dfrac{k_r}{2} \varphi_1^2+\dfrac{E A}{2 l} \left[(l'_{+} - l )^2+(l'_{-} - l )^2\right] + \Upsilon_{ic} - F u ,
\end{equation}
 can be found by imposing $\partial\Pi/\partial\varphi_1=\partial\Pi/\partial\varphi_2=\partial\Pi/\partial\Delta=0$. In particular, the work done by the force $F$ can be calculated by expressing the magnitude of the horizontal displacement at the right end as
\begin{equation}
\label{uu}
u= \vert \left(R+L/2\right) \cos\varphi_2- \left(R-L/2-\Delta\right) \cos\varphi_1 - L \vert ,
\end{equation}
where $L/2$ is the length of each of the two rigid tracts, so that $L$ is the whole length of the undeformed system, and $R$ is the radius of curvature of the joint, with $R>L/2$ (weak curvature) in the considered anatomical plane. Moreover, the internal elastic energy is provided by the sum of those stored during the deformation inside of the rotational spring at the hinge and of the two bands, both modelled as linear elastic springs, plus the energetic contribution $\Upsilon_{ic}$. The latter is due to the suction effect resisting distraction, which leads to the development of an intra-capsular pulling back pressure in response to increases of the articular volume. More in detail, $k_r$ is the stiffness of the elastic hinge, while $E$, $A$ and $l$ are the Young modulus, the nominal cross-sectional area and the length at rest of the bands, respectively, with $l$ accounting for the effective length covered by ligaments in real fingers' articulation. The deformed lengths of the lower and upper bands are 
$l'_{+}$ and $l'_{-}$, respectively, and can be calculated as
\begin{equation}
\label{lpm}
l'_{\pm} =  l \, + \left\lbrace2R^2\left[1-\cos\left(\varphi_1-\varphi_2\right)\right]+ \Delta \left\lbrace \Delta -2R \left[\cos\gamma+\cos\left(\gamma \pm \varphi_1 \mp \varphi_2\right)\right]\right\rbrace \right\rbrace^{1/2},
\end{equation}
where $\gamma$ is the opening angle of the joint, $\gamma = \arcsin \left( h/2R\right)$, $h$ being its transverse size. Finally, the energy  $\Upsilon_{ic}$ can be expressed as
\begin{equation}
\Upsilon_{ic}= \dfrac{V_{ic} K_{ic}}{\beta}\left[e^{\beta \left(J_{ic}-1\right)}-1-\beta\left(J_{ic}-1\right)\right],
\end{equation}
obtained by assuming the standard exponential law
\begin{equation}
\label{pic}
p_{ic}=\dfrac{\partial\Upsilon_{ic}}{\partial J_{ic}}=K_{ic}\left[e^{\beta\left(J_{ic}-1\right)}-1\right]
\end{equation} 
to describe how the intra-capsular pressure $p_{ic}$ varies with the Jacobian of the deformation $J_{ic}=V_{ic}'/V_{ic}$, representing the ratio between the volumes of the articular fluid-filled space in the deformed and undeformed states, approximated as being spatially homogeneous. Here, $K_{ic}$ is the overall bulk modulus of the synovial fluid and matter inside the intra-articular space and $\beta$ is a constitutive parameter ruling the grow rate of the pressure. Moreover, it is possible to consider $V_{ic}=s A_{ic}$ and $V'_{ic}= s A'_{ic}$, where $s$ identifies the out-of-plane size of the joint  (remaining invariant during the deformation process), while $A_{ic}$ and $A'_{ic}$ are the areas occupied by the projection of the intra-capsular domain onto the horizontal anatomical plane at the reference and current configurations of the system, respectively. As sketched in Fig. \ref{fig.Figuretc}, the reference area can be estimated as $A_{ic} \approx h t$, where $t$ is a nominal thickness, while geometrical considerations allow to obtain the following form for the current area $A_{ic}'$:
\begin{equation}
A_{ic}' =A_{ic}+R \Delta \left[1+\cos\left(\varphi_1-\varphi_2\right)\right]\sin\gamma.
\end{equation}
The above equations led to a numerical solution, with the aid of the software Mathematica\textsuperscript{\textregistered}. The diagrams reported in Fig. \ref{fig.Figuretc} (red curves) show how, due to the cooperation of the elasticity of the ligaments and the suction associated to the intra-capsular pressure, the bifurcation tensile load $F_{cr}^{+}$ is  attained at the end of an initial deformation process connected with the axial distraction of the bone segments and is followed by a significant reduction in the force gradient with respect to the end displacement, when the post-buckling phase evolves. This mechanical behaviour demonstrates how, in case of high distracting loads, tensile buckling rules lateral luxation and provides an overall stress shielding by deviating fingers from their natural straight configuration.

\subsection{\textit{Dorsal/volar} luxations ruled by compressive buckling along the strong curvature}
Dorsal and volar luxations of fingers are injuries complementary to lateral ones and occur in response to high compressive actions. In this case, the digits dislocate in the sagittal plane of the hand, where articular joints exhibit strong curvatures. From a mechanical point of view, these injuries can again be interpreted as elastic instabilities, which lead the bone-joint-bone system to deviate its configuration at the onset of compressive buckling as shown on the right of Fig. \ref{fig.Figuretc}. In more detail, the mechanics of dorsal/volar luxations can be treated by following an approach analogous to that employed for lateral ones. However, in this case, the rotations $\varphi_1$ and $\varphi_2$ are the sole Lagrangian variables needed for a complete description of the kinematics of the system, as no separation at the joint interface occurs under compression. As a consequence, the total potential energy \eqref{PiT}  lacks now the energy term $\Upsilon_{ic}$, and the expressions for the displacement $u$ in \eqref{uu} and the lengths of the two elastic ligaments $l'_{\pm}$ in \eqref{lpm} simplify by imposing  $\Delta=0$.\\
Stationarity of $\Pi$ allows to find the equilibrium bifurcation diagrams shown in Fig. \ref{fig.Figuretc} (blue curves) and the critical compressive load can be obtained in closed-form as
\begin{equation}
\label{Fcrcc}
F_{cr}^{-}=\dfrac{\sqrt{a_1^2+4 a_0 a_2}-a_1}{2 a_0},
\end{equation} 
the coefficients $a_0$, $a_1$ and $a_2$ being:
\begin{equation}
\begin{split}
a_0 &=l \left(4 R^2-L^2\right), \\
a_1 &=2 \left[k_r l \left(2R+L\right)+4 E A R^2 L\right], \\
a_2 &= 8 E A k_r R^2,
\end{split}
\end{equation}
with $R<L/2$ (strong curvature) and $a_1^2+4a_0 a_2 \geq 0$.
It is worth noticing that the critical value \eqref{Fcrcc} corresponds to the minimum between two possible bifurcation loads.\\
Similarly to the lateral luxation (governed by tensile buckling), the mechanical  model demonstrates that dorsal/volar dislocations, ruled by compressive elastic instability, provide again stress shielding of the bone-joint-bone finger system in the sagittal plane where digits are deviated, invited by the strong curvature of the articulating bone segments' heads (see bifurcation diagrams in Fig. \ref{fig.Figuretc}).

\subsection{An experimental proof of concept prototype}
The proposed luxation mechanism is validated by an experimental proof of concept model, based on a 3D-printed prototype, incorporating the suction effect due to the articular capsule and a design of the elastic ligaments tailored to replicate the real functioning of the fingers' bone-joint-bone system, as reported in Fig. \ref{fig.Figure3}. Laboratory uni-axial tests performed on the built up prototype has  confirmed that both compressive and tensile buckling modes can be reproduced if all the main geometrical and mechanical features characterizing the articulation are taken into account. In particular, under compression, instability-induced deviation of the system's tracts from the straight configuration  occurs in the strong curvature plane, as it happens for dorsal and volar dislocations, while bifurcation associated to tension takes place in the weak curvature plane, after an initial phase of axial distraction, similarly to lateral dislocations. These evidences support the theoretical model and contribute to experimentally highlight the double capability of finger joints to undergo both compressive and tensile buckling, so  providing luxation as a stress shielding strategy.    

\section{Discussion and conclusion}
The main conclusion of the present study is that 
dislocations --common injuries of bones articulations as finger joints-- can be interpreted through an unprecedented mechanical instability phenomenon that unifies  compressive and tensile buckling and leads to recognise luxation as a protection mechanism from more severe damage for bones, exposed to extreme load. This is related to the characteristic double-curvature geometry of the bones' epiphyses at the joints, allowing for a two-mode (lateral and dorsal/volar) fingers' dislocation model, which has  theoretically and experimentally demonstrated  to be the result of a unique optimized design by nature.  The latter ensures stress shielding under axial forces exceeding physiological limits, independently from their sign. In tension, the proposed mechanism provides the first example of tensile critical load involved in functioning of a natural system.\\
The outcomes of the present research can be used in the design of new bio-inspired, mechanically and geometrically optimized joints and structures, which can find a number of applications at different scales, from soft- and micro-robotics to active and passive actuators and exoskeletons in healthcare for rehabilitation purposes.

\begin{figure*}[htbp] %htbp
\centering
\includegraphics[width=.89\textwidth]{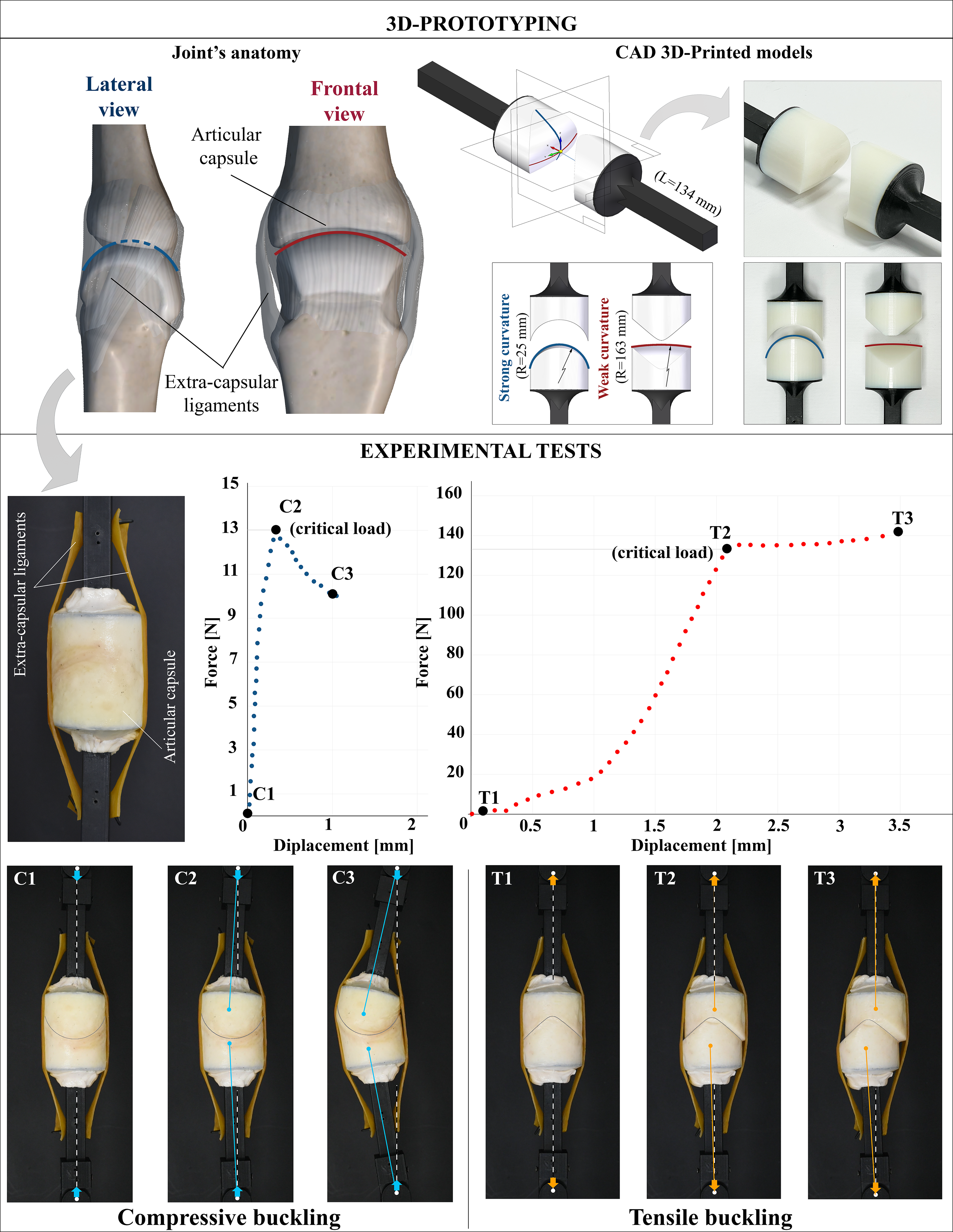}
\caption{\textbf{Upper part:} Lateral and frontal views of the fingers' joint anatomy pointing out the two characteristic (strong and weak) curvatures of the articulating bone terminals (left) are compared with the CAD model and 3D-printed prototype of the finger bone-joint-bone system 
(right). \textbf{Central part:} Assembled prototype employed for experimental tests --equipped with an articular capsule obtained from sow intestine, lubricating grease at the joint interface and ligament-like elastic bands-- (left) produces the compressive (blue) and tensile (red) force-displacement curves  (right). 
\textbf{Lower part:} Sequences of images showing the buckling kinematics exhibited by the system during the compressive (frames C1-C3) and tensile (frames T1-T3) tests; in particular, frames C1 and T1 correspond to initial (unloaded and undeformed) configurations, frames C2 and T2 show states immediately following the onset of equilibrium bifurcation and frames C3 and T3 provide configurations reached in the post-buckling phase. It is worth noticing that, due to the partial compressibility of the grease layer at the interface, the response in compression does not recover the infinitely rigid  behaviour theoretically predicted. Also, consistently with the anatomy of the finger joints and the theoretical model, elastic bands were positioned laterally in the sagittal and frontal planes and are shown in the photos deformed under compression or tension.}
\label{fig.Figure3}
\end{figure*}

\paragraph*{Materials and Methods}
The proof of principle  prototype of the fingers' bone-joint-bone system was manufactured through 3D printing and tested under both tensile and compressive forces. A double-curvature joint was designed to replicate (scale 3:1) the two complementary curvatures exhibited by the finger bones' epiphyses 
in a 3D printed model realized with STRATASYS F170 and STRATASYS Object30 printer, using polymeric materials (ABS - Acrylonitrile Butadiene Styrene) for the bone tracts and a resinous material for the joint elements, Fig. \ref{fig.Figure3}. Particular attention was devoted to mimic both the lubrication related to the synovial fluid and the suction effect occurring within the articular capsule. To the purpose, a film of high-pressure resistant grease (NLGI 3 lithium by AREXONS) was interposed between the two contact surfaces of the joined elements and an intact portion of sow intestine was used to encapsulate the joint. Finally, elastic bands were fixed on the sides of the assembled structure in order to simulate the presence of extra-capsular ligaments, as illustrated in Fig. \ref{fig.Figure3}. The mechanical response of the prototype was measured  using a electromechanical bi-axial testing machine  (ElectroForce TestBench four linear motors Planar Biaxial $230 V$ with integrated $200 N$ load-cells by TA Instruments) used in an uniaxial mode to reproduce both tensile and compressive loading conditions. Specifically, a progressively increasing axial displacement was quasi-statically (at a velocity of $0.05$~mm/s) applied to the ends of the structure, both constrained through 3D printed hinges. The reaction forces on the supports were measured via the above specified $200 N$ load cells.

%\paragraph{Ethics.}The authors declare that all the ethical issues were respected.

\paragraph{Data Accessibility.} All study data are included in the main text.

\paragraph{Authors’ Contributions.}All the authors equally contributed to the work.

\paragraph{Competing Interests.}The authors declare no competing interest.

\paragraph{Acknowledgements.}All the authors acknowledge the financial support from the European Research Council (ERC) under the European Union’s Horizon 2020 research and innovation programme, Grant agreement No. ERC-ADG-2021-101052956-BEYOND. MF, SP, AC, and ARC also acknowledge the support by the Italian Ministry of University and Research (MUR), under the complementary actions to the NRRP "Fit4MedRob - Fit for Medical Robotics" Grant (\# PNC0000007).

\newpage

\bibliographystyle{ieeetr}
\bibliography{Bib.bib}

\begin{thebibliography}{10}

\bibitem{graysanatomy}
H.~Gray, {\em Anatomy of the human body. 20th Ed.}
\newblock Philadelphia Lea and Febiger, 1918.

\bibitem{guidugli}
M.~Brocato and P.~Podio-Guidugli, ``A bone-wise approach for modeling the human hind-midfoot. part i: kinematics,'' {\em Meccanica}, vol.~57, no.~5, pp.~977--998, 2022.

\bibitem{Gray}
J.~Gray, {\em How {A}nimals {M}ove}.
\newblock Cambridge University Press, 1953.

\bibitem{ralphs1994}
J.~Ralphs and M.~Benjamin, ``The joint capsule: structure, composition, ageing and disease.,'' {\em Journal of anatomy}, vol.~184, no.~Pt 3, p.~503, 1994.

\bibitem{Mow1993}
V.~C. Mow, G.~A. Ateshian, and R.~L. Spilker, ``{Biomechanics of Diarthrodial Joints: A Review of Twenty Years of Progress},'' {\em Journal of Biomechanical Engineering}, vol.~115, pp.~460--467, 11 1993.

\bibitem{neville2007}
A.~Neville, A.~Morina, T.~Liskiewicz, and Y.~Yan, ``Synovial joint lubrication—does nature teach more effective engineering lubrication strategies?,'' {\em Proceedings of the Institution of Mechanical Engineers, Part C: Journal of Mechanical Engineering Science}, vol.~221, no.~10, pp.~1223--1230, 2007.

\bibitem{tissuemechanics}
S.~C. Cowin and S.~B. Doty, {\em Tissue mechanics}.
\newblock Springer, 2007.

\bibitem{fungbook}
Y.-c. Fung, {\em Biomechanics: mechanical properties of living tissues}.
\newblock Springer Science \& Business Media, 2013.

\bibitem{NIPhand1}
W.-C. Hung, C.-H. Chang, A.-T. Hsu, and H.-T. Lin, ``The role of negative intra-articular pressure in stabilizing the metacarpophalangeal joint,'' {\em Journal of Mechanics in Medicine and Biology}, vol.~13, no.~02, p.~1350049, 2013.

\bibitem{NIPshoulder1}
W.~Inokuchi, B.~S. Olsen, J.~O. S{\o}jbjerg, and O.~Sneppen, ``The relation between the position of the glenohumeral joint and the intraarticular pressure: an experimental study,'' {\em Journal of Shoulder and Elbow Surgery}, vol.~6, no.~2, pp.~144--149, 1997.

\bibitem{tribon}
G.~N. Kawchuk, J.~Fryer, J.~L. Jaremko, H.~Zeng, L.~Rowe, and R.~Thompson, ``Real-time visualization of joint cavitation,'' {\em PloS one}, vol.~10, no.~4, p.~e0119470, 2015.

\bibitem{NIPshoulder2}
P.~Habermeyer, U.~Schuller, and E.~Wiedemann, ``The intra-articular pressure of the shoulder: an experimental study on the role of the glenoid labrum in stabilizing the joint,'' {\em Arthroscopy: The Journal of Arthroscopic \& Related Surgery}, vol.~8, no.~2, pp.~166--172, 1992.

\bibitem{shoulderstab1}
T.~D. Gibb, J.~A. Sidles, D.~Harryman~2nd, K.~J. McQuade, and F.~Matsen~3rd, ``The effect of capsular venting on glenohumeral laxity.,'' {\em Clinical orthopaedics and related research}, no.~268, pp.~120--127, 1991.

\bibitem{NIPshoulder3}
S.~Alexander, D.~F. Southgate, A.~M. Bull, and A.~L. Wallace, ``The role of negative intraarticular pressure and the long head of biceps tendon on passive stability of the glenohumeral joint,'' {\em Journal of shoulder and elbow surgery}, vol.~22, no.~1, pp.~94--101, 2013.

\bibitem{fingerstab}
N.~Sharma and M.~Venkadesan, ``Finger stability in precision grips,'' {\em Proceedings of the National Academy of Sciences}, vol.~119, no.~12, p.~e2122903119, 2022.

\bibitem{PIPjoints}
D.~Ramponi and M.~J. Cerepani, ``Finger proximal interphalangeal joint dislocation,'' {\em Advanced emergency nursing journal}, vol.~37, no.~4, pp.~252--257, 2015.

\bibitem{fingerdisloc}
R.~B. Prucz and J.~B. Friedrich, ``Finger joint injuries,'' {\em Clinics in Sports Medicine}, vol.~34, no.~1, pp.~99--116, 2015.
\newblock Sports Hand and Wrist Injuries.

\bibitem{SUNDARAM2013}
N.~Sundaram, J.~Bosley, and G.~S. Stacy, ``Conventional radiographic evaluation of athletic injuries to the hand,'' {\em Radiologic Clinics of North America}, vol.~51, no.~2, pp.~239--255, 2013.
\newblock Imaging of Athletic Injuries of the Upper Extremity.

\bibitem{elzinga2017finger}
K.~E. Elzinga and K.~C. Chung, ``Finger injuries in football and rugby,'' {\em Hand clinics}, vol.~33, no.~1, pp.~149--160, 2017.

\bibitem{basketball}
L.~Laver, B.~Kocaoglu, B.~Cole, A.~J. Arundale, J.~Bytomski, and A.~Amendola, {\em Basketball Sports Medicine and Science}.
\newblock Springer, 2020.

\bibitem{bach1999finger}
A.~W. Bach, ``Finger joint injuries in active patients: pointers for acute and late-phase management,'' {\em The Physician and Sportsmedicine}, vol.~27, no.~3, pp.~89--104, 1999.

\bibitem{logan2004}
A.~J. Logan, N.~Makwana, G.~Mason, and J.~Dias, ``Acute hand and wrist injuries in experienced rock climbers,'' {\em British journal of sports medicine}, vol.~38, no.~5, pp.~545--548, 2004.

\bibitem{gnecchi2015}
S.~Gnecchi, F.~Moutet, {\em et~al.}, ``Hand and finger injuries in rock climbers,'' tech. rep., Springer, 2015.

\bibitem{zaccaria2011}
D.~Zaccaria, D.~Bigoni, G.~Noselli, and D.~Misseroni, ``Structures buckling under tensile dead load,'' {\em Proceedings of the Royal Society A: Mathematical, Physical and Engineering Sciences}, vol.~467, no.~2130, pp.~1686--1700, 2011.

\bibitem{elasticstability}
S.~P. Timoshenko and J.~M. Gere, {\em Theory of elastic stability}.
\newblock McGraw-Hill, 1961.

\bibitem{Bigoni2012}
D.~Bigoni, {\em Nonlinear Solid Mechanics - Bifurcation theory and material instability}.
\newblock 2012.

\bibitem{1dsd}
S.~Palumbo, L.~Deseri, D.~R. Owen, and M.~Fraldi, ``Disarrangements and instabilities in augmented one-dimensional hyperelasticity,'' {\em Proceedings of the Royal Society A: Mathematical, Physical and Engineering Sciences}, vol.~474, no.~2218, p.~20180312, 2018.

\bibitem{paradox}
M.~Fraldi, S.~Palumbo, A.~Cutolo, A.~R. Carotenuto, and F.~Guarracino, ``On the equilibrium bifurcation of axially deformable holonomic systems: solution of a long-standing enigma,'' {\em Proceedings of the Royal Society A: Mathematical, Physical and Engineering Sciences}, vol.~477, no.~2253, p.~20210327, 2021.

\bibitem{siviy2022}
C.~Siviy, L.~M. Baker, B.~T. Quinlivan, F.~Porciuncula, K.~Swaminathan, L.~N. Awad, and C.~J. Walsh, ``Opportunities and challenges in the development of exoskeletons for locomotor assistance,'' {\em Nature Biomedical Engineering}, pp.~1--17, 2022.

\bibitem{Misseroni_2015}
D.~Misseroni, G.~Noselli, D.~Zaccaria, and D.~Bigoni, ``The deformation of an elastic rod with a clamp sliding along a smooth and curved profile,'' {\em International Journal of Solids and Structures}, vol.~69-70, pp.~491--497, sep 2015.

\bibitem{hu2014}
D.~Hu, D.~Howard, and L.~Ren, ``Biomechanical analysis of the human finger extensor mechanism during isometric pressing,'' {\em PloS one}, vol.~9, no.~4, p.~e94533, 2014.

\bibitem{ventose}
D.~Bigoni, N.~Bordignon, A.~Piccolroaz, and S.~Stupkiewicz, ``Bifurcation of elastic solids with sliding interfaces,'' {\em Proceedings of the Royal Society A: Mathematical, Physical and Engineering Sciences}, vol.~474, no.~2209, p.~20170681, 2018.

\end{thebibliography}

\end{document}